\documentclass[12pt]{article}
\usepackage[centertags]{amsmath}
\usepackage{setspace}
\usepackage{amsfonts}
\usepackage{amssymb}
\usepackage{amsthm}
\usepackage{cancel}
\usepackage{color}
\newcommand{\beq}{\begin{equation}}
\newcommand{\eeq}[1]{\label{#1}\end{equation}}
\newcommand{\bea}{\begin{eqnarray}}
\newcommand{\eea}[1]{\label{#1}\end{eqnarray}}

\def\de{\partial}
\def\nb{\nabla}
\def\trr{\triangleright}

 \csname
@addtoreset\endcsname{equation}{section}

\def\a{\alpha}
\def\b{\beta}
\def\g{\gamma}

\def\d{\delta}
\def\D{\Delta}
\def\e{\epsilon}
\def\ve{\varepsilon}

\def\h{\eta}

\def\l{\lambda}
\def\L{\Lambda}
\def\m{\mu}
\def\n{\nu}

\def\r{\rho}

\def\s{\sigma}
\def\S{\Sigma}

\def\vf{\varphi}

\begin{document}
\setlength{\topmargin}{-1cm} \setlength{\oddsidemargin}{0cm}
\setlength{\evensidemargin}{0cm}
\begin{titlepage}
\begin{center}
{\Large \bf Consistent Non-Minimal Couplings of\\ Massive Higher-Spin Particles}

\vspace{20pt}

{\large Ignacio Cortese$^a$, Rakibur Rahman$^a$ and M. Sivakumar$^b$}

\vspace{12pt}
$a$) Physique Th\'eorique et Math\'ematique \& International Solvay Institutes\\
     Universit\'e Libre de Bruxelles, Campus Plaine C.P. 231, B-1050 Bruxelles, Belgium\\
\vspace{6pt}
$b$) School of Physics, University of Hyderabad, Hyderabad 500046, India\\
\vspace{6pt}
e-mail: {\small \it
{icortese@ulb.ac.be, rakibur.rahman@ulb.ac.be, mssp@uohyd.ernet.in}}\vspace{10pt}

\end{center}

\vspace{20pt}

\begin{abstract}
The mutual compatibility of the dynamical equations and constraints describing a massive particle of arbitrary
spin, though essential for consistency, is generically lost in the presence of interactions. The conventional Lagrangian
approach avoids this difficulty, but fails to ensure light-cone propagation and becomes very cumbersome. In this
paper, we take an alternative route$-$the involutive form of the equations and constraints$-$to guarantee their
algebraic consistency. This approach enormously simplifies the search for consistent interactions, now seen as
deformations of the involutive system, by keeping manifest the causal propagation of the correct number of
degrees of freedom. We consider massive particles of arbitrary integer spin in electromagnetic and gravitational
backgrounds to find their possible non-minimal local couplings. Apart from easily reproducing some well-known
results, we find restrictions on the backgrounds for consistent propagation of such a particle in isolation.
The results can be altered by non-local interactions that may arise from additional massive states in the
interacting theory.

\end{abstract}

\end{titlepage}

\newpage
\section{Introduction}

Any fundamental particle described in Quantum Field Theory carries an irreducible unitary representation of the Poincar\'{e}
group. Massive particles of arbitrary spin, which belong to the first Wigner class, are customarily represented by symmetric
traceless tensors (bosons) or symmetric $\gamma$-traceless tensor-spinors (fermions)\footnote{While spin is not a good quantum
number in $d>4$, the rank of the symmetric ($\gamma$-)traceless Lorentz tensor(-spinor) continues to define ``spin'' in
arbitrary dimensions.}. A spin-$s$ bosonic field of mass $m$, which we denote by $\vf_{\m_1...\m_s}$, is required to satisfy
the Klein-Gordon equation, \beq (\de^2-m^2)\vf_{\m_1...\m_s}=0,\eeq{int1} and is subject to the divergence condition,
\beq \de\cdot\vf_{\m_1...\m_{s-1}}\equiv\de^{\m_s}\vf_{\m_1...\m_s}=0.\eeq{int2} Of course, the field $\vf_{\m_1...\m_s}$
is traceless to begin with: \beq \vf'_{\m_1...\m_{s-2}}\equiv\vf_{\m_1...\m_{s-1}}^{~~~~~~~~~\m_{s-1}}=0.\eeq{int3}
The dynamical equation~(\ref{int1}) and the constraints~(\ref{int2}) and~(\ref{int3}) comprise a set of Fierz-Pauli conditions,
from which one finds that in $d$ dimensions the total number of propagating degrees of freedom (DoF) is given by
\beq \mathfrak{D}={d-4+s\choose s}+2\,{d-4+s\choose s-1}.\eeq{int4}
In particular when $d=4$, this number reduces to $2s+1$ as expected.

As first noted by Fierz and Pauli~\cite{pf}, turning on interactions for these higher-spin (HS) fields at the level of
equations of motion (EoM) and constraints, by replacing ordinary derivatives with covariant ones in
Eqs.~(\ref{int1})-(\ref{int3}), results in inconsistencies. Consider, for example, a massive spin-$s$ field,
$\vf_{\m_1...\m_s}$, minimally coupled to electromagnetism (EM). The na\"ive covariantization,
$\de_\m\rightarrow D_\m=\de_\m+ieA_\m$, of Eqs.~(\ref{int1})-(\ref{int3}) gives
\beq \left(D^2-m^2\right)\vf_{\m_1...\m_s}=0,\qquad D\cdot\vf_{\m_1...\m_{s-1}}=0,\qquad \vf'_{\m_1...\m_{s-2}}=0.\eeq{int5}
The Klein-Gordon equation and the transversality condition, however, yield
\beq \left[D^{\m_1},D^2-m^2\right]\vf_{\m_1...\m_s}=0,\eeq{int6} which results in unwarranted constraints because covariant
derivatives do not commute. For a constant EM field strength $F_{\m\n}$, for example, one gets
\beq ieF^{\m_1\r}D_\r\vf_{\m_1...\m_s}=0.\eeq{int7} This constraint disappears when the interaction is turned off, and so
the system~(\ref{int5}) does not describe the same number of DoFs as the free theory. To avoid such difficulties, Fierz and Pauli
suggested~\cite{pf} that one take recourse to the Lagrangian formulation, which would automatically render the resulting EoMs and
constraints algebraically consistent.

However, a Lagrangian formulation guarantees neither that no unphysical DoFs start propagating nor that the physical ones propagate
only within the light cone. Indeed, superluminal propagation can occur in non-trivial external EM backgrounds even for infinitesimally
small values of the EM field invariants. This is the notorious Velo-Zwanziger problem~\cite{vz}. This pathology manifests itself in
general for all charged massive HS particles with $s\geq3/2$. Field theoretically it is quite challenging to construct consistent
interactions of massive HS particles since this problem persists for a wide class of non-minimal generalizations of the
theory and also for other interactions~\cite{sham,kob,d3}.

Addition of non-minimal terms and/or new dynamical DoFs may rescue causality. For a massive charged spin-$\tfrac{3}{2}$ field,
the problem is elegantly solved by $\mathcal{N}=2$ (broken) supergravity~\cite{SUGRA1,SUGRA2,DZ} or by judiciously constructed
non-minimal models~\cite{PR2}. For $s\geq2$, the only explicit solution known to date comes from open string field
theory~\cite{AN2,PRS}, which spells out highly non-minimal terms so that any field belonging to the first Regge trajectory
propagates causally in a constant EM background. Explicit string-theoretic Lagrangians are known for $s=2$~\cite{AN2} and
$s=3$~\cite{string3}, and they are guaranteed to exist for any HS field~\cite{PRS}. These horribly complicated Lagrangians
give rise in the critical dimension a very simple but consistent set of Fierz-Pauli conditions~\cite{AN2,PRS}:
\beq \left(D^2-m^2\right)\vf_{\m_1...\m_s}-2iesF^\a_{~(\mu_1}\vf_{\mu_2...\mu_s)\a}=0,\qquad D\cdot\vf_{\mu_1...\mu_{s-1}}=0,
\qquad\vf'_{\mu_1...\mu_{s-2}}=0.\eeq{String-EM} The enormous simplicity at the level of EoMs and constraints makes one
wonder whether a Lagrangian formulation, originally proposed in~\cite{pf}, is really the best way of understanding
HS interactions. After all, in the context of massless HS fields consistent interacting theories appear in AdS space at the
level of EoMs~\cite{Vasiliev}, which have resisted so far any embedding into a Lagrangian framework, if it exists at all.

Given this, one can step back to revisit the issue of introducing interactions at the level of the EoMs and constraints.
Notice that the free system~(\ref{int1})-(\ref{int3}) and its consistent deformation~(\ref{String-EM}) are strikingly similar:
they both not only set the divergence and trace exactly to zero, but also have in common some not-so-apparent features that are
important for consistency\footnote{We will investigate and exploit these features, \`a la Ref.~\cite{KaLySh3}, to construct
consistent interactions.}. While the na\"ive covariantization~(\ref{int5}) fails, the consistent set~(\ref{String-EM}) is not
a major modification either. Is it possible to find a systematic procedure to deform
the free equations in the presence of interactions without hurting their algebraic consistency?
The answer is yes. Indeed, the authors of Ref.~\cite{KaLySh3} have addressed precisely this issue and proposed a universal
covariant method for constructing consistent interactions at the level of field equations. Unlike the Lagrangian framework,
their method may not need any auxiliary fields and relies on the involution and preservation of gauge symmetries and identities
(to be explained in Section~\ref{sec:Formalism}) of the EoMs and constraints, which guarantee algebraic consistency. This approach
may simplify the search for consistent interactions to a great extent by keeping manifest the causal propagation of the correct
number of DoFs. In this paper, we employ this method to find non-minimal local couplings of massive particles of arbitrary
integer spin exposed to external EM and gravitational backgrounds.

The organization of the paper is as follows. In the remaining of this Section we clarify our conventions and notations and
present our main results. In Section~\ref{sec:Formalism} we get familiar with the formalism proposed in Ref.~\cite{KaLySh3}
by explaining some key ideas and working out warm-up examples of a massive spin-1 particle in EM and gravitational backgrounds.
Section~\ref{sec:EMs} is devoted to both EM and gravitational interactions of massive particles of arbitrary
integer spin. A general methodology is developed throughout this Section, which we apply in either case to find the possible
non-minimal local couplings and identify the backgrounds that may consistently propagate such a particle in isolation.
The resulting couplings are the magnetic dipole and the gravitational quadrupole moments, quantified respectively by the
$g$- and the $h$-factors. Their values are examined in Section~\ref{sec:ghNL}, where we also see how they may get modified by
the presence of non-locality and/or additional dynamical fields. We conclude in Section~\ref{sec:Conclusion} with some remarks.

\subsubsection*{Conventions \& Notations}

We work with a mostly positive metric. The notation $(i_1\cdots i_n)$ means totally symmetric expression in all the indices
$i_1,\dots,i_n$ with the normalization factor $\tfrac{1}{n!}$. The tensor $\eta^{\a_1\dots\a_n,\m_1\dots\m_n}\equiv\eta^{\a_1\b_1}\cdots\eta^{\a_n\b_n}\delta^{\m_1\dots\m_n}_{\b_1\dots\b_n}$
will appear in many places.
We denote EM and gravitational covariant derivatives respectively as $D_\m$ and $\nb_\m$, whose commutators obey
\bea &[D_\m,D_\n]=ieF_{\m\n},&\nonumber\\&[\nb_\m,\nb_\n]V^\r=R^\r_{~\s\m\n}V^\s.&\nonumber\eea{commutators}

\subsubsection*{Results}

\begin{itemize}
 \item In isolation a massive charged HS particle with local EM interactions may consistently propagate only as a probe
 in an EM background. For $s=1$, the background is required to satisfy the source-free Maxwell equations:
 $\de_\m F^{\m\n}=0$, whereas for $s\geq2$, the symmetrized gradient of the field strength must vanish: $\de_{(\m}F_{\n)\r}=0$.

 \item Consistent local gravitational interactions of a solitary massive HS particle may exist only in an external
 gravitational background. The Ricci tensor of this manifold must be covariantly constant (Ricci symmetric space), and
 for $s\geq 3$, the gradient of the Weyl tensor must also satisfy: $\nabla_{(\m}W_\n{}^\a{}_{\r)}{}^\b=0$. No such
 restrictions exist for $s=1$.

 \item The covariant transversality condition in both cases requires no modification.

 \item The above results$-$derived for irreducible representations: symmetric Lorentz tensors with vanishing trace$-$hold
 whether or not the system comes from a Lagrangian, provided that interactions are local and no other DoFs are present.
 Therefore, consistent propagation of massive HS particles in arbitrary EM and gravitational backgrounds calls for
 non-locality and/or a (possibly infinite) tower of massive states.

 \item The gyromagnetic ratio or $g$-factor, that quantifies the magnetic dipole moment of the particle, must be $g=2$. Other
 values are possible only when non-local interactions and/or additional massive states are present.

 \item The gravimagnetic ratio or $h$-factor of the particle, that quantifies its gravitational quadrupole moment, must be $h=1$.
 This value may get altered again by non-local interactions and/or the presence of other massive particles.
 \end{itemize}

\section{Formalism \& Warm-Up Examples}\label{sec:Formalism}

In this Section, we explain some basic notions to summarize the formalism$-$the deformation of involutive
equations$-$proposed in Ref.~\cite{KaLySh3} for covariant construction of consistent interactions. We will skip some
technical details; readers may take a look at Ref.~\cite{KaLySh3} and references therein. While our ultimate
goal is to study the EM and gravitational couplings of massive particles of any integer spin, we will consider all along
the example of spin 1 for the sake of simplicity. The methodology for arbitrary spin will be developed in
Section~\ref{sec:EMs}, which will then be employed to find consistent non-minimal couplings.

\subsubsection*{Involution}

Let us consider a system of partial differential equations
\beq T^a[\Phi^i,\de_\m\Phi^i,\dots,\de_{\m_1}\cdots\de_{\m_q}\Phi^i]=0,\qquad a=1,2,...,t,\eeq{eq:eom}
that governs the dynamics of some fields $\Phi^i$, $i=1,2,...,f$. The maximal order of these equations defines
the order of this system, which is $q$. The system~(\ref{eq:eom}) is said to be involutive if it contains
all the differential consequences of $\text{order}\leq p$ derivable from any order-$p$ subsystem:
$T^b[\Phi^i,\de_\m\Phi^i,\dots,\de_{\m_1}\cdots\de_{\m_p}\Phi^i]=0,\,b\subset a,\,p\leq q$.

To illustrate the difference between involutive and non-involutive systems, let us consider the second order Lagrangian
EoMs of the Proca field $\vf^\m$ in flat space-time
\beq \de^2\vf_\m-\de_\m\de\cdot\vf-m^2\vf_\m=0,\qquad m^2\neq0.\eeq{man1}
Its divergence however gives rise to the first order transversality condition:
\beq \de\cdot\vf=0.\eeq{man2} The Proca system~(\ref{man1}) does not include this lower order differential consequence,
and is therefore non-involutive. On the other hand, when the EoMs~(\ref{man1}) are supplemented by Eq.~(\ref{man2}), they
leave us with a second-order involutive system of equations: \beq T^\m=\left(\de^2-m^2\right)\vf^\m=0,\qquad
T=\de\cdot\vf=0,\eeq{eq:involutiveeom} which is of course equivalent to the original non-involutive one~(\ref{man1}).
The involutive system~(\ref{eq:involutiveeom}) is non-Lagrangian as it consists of $d+1$ equations$-$too
many to result directly from the variation of a Lagrangian functional of the $d$-component field $\vf_{\m}$.

As a matter of fact, any field theory can be brought to an involutive form~\cite{KaLySh3}, which is equivalent to the
original system in that they both have the same solution space. Generically, an involutive system may or may not be
a Lagrangian one. Free massive HS fields, in particular, have non-involutive Lagrangian equations, but they can also be
described by an involutive non-Lagrangian system, namely Eqs.~(\ref{int1})-(\ref{int3}). The involutive form retains
all the symmetries of the original system, and can be very useful in the study of covariant field equations, as we will see.

\subsubsection*{Gauge Symmetries \& Gauge Identities}

Let the system~(\ref{eq:eom}) be involutive. In general, it may enjoy local gauge symmetries
\beq \delta_\ve\Phi^i=\ve^\a R_\a^i,\qquad\delta_\ve T_a|_{\,T=0}=0,\qquad \a=1,2,\dots,r,\eeq{gaugesymm}
where $\ve^\a$ are the gauge parameters, while $R_\a^i$ are the gauge symmetry generators, which are differential
operators of finite order for local symmetries.

More importantly, the involutive system may possess non-trivial gauge identities, which may or may not be
related to gauge symmetries. Their schematic form is
\beq L^{A}\trr T\equiv L_a^A T^a=0,\qquad A=1,2,\dots,l,\eeq{eq:lgen}
with the gauge identity generators $L^A_a$ being local differential operators. Note that for Lagrangian systems there
is an isomorphism between symmetries and Noether identities. For generic non-Lagrangian involutive systems, no such
correspondence exists; still one can have non-trivial gauge identities. In other words, gauge identities are more
generic than Noether ones, and may exist even in the absence of any gauge symmetry. However, the two coincide for a
set of Lagrangian equations if they are involutive from the outset~\cite{KaLySh3}.

Our spin-1 example do not have any gauge symmetry, but it is easy to see that the involutive form~(\ref{eq:involutiveeom})
possesses the gauge identity \beq L\trr T\equiv L^\m T_\m+LT=\de^\m T_\m-\left(\de^2-m^2\right)T=0,\eeq{gaugeid01} where
the gauge identity generators are given by \beq L^\m=\de^\m,\qquad L=-\left(\de^2-m^2\right).\eeq{eq:gidgens1}
Eq.~(\ref{gaugeid01}) is a third order gauge identity. In general, the order of the gauge identity~(\ref{eq:lgen})
is defined as the maximal order of individual terms appearing in the summation $L_a^A T^a$.

The gauge identities of an involutive system are important in that they reflect algebraic consistency and play a crucial
role in the DoF count, to which we now turn.

\subsubsection*{Compatibility \& DoF Count}

The compatibility coefficient of an involutive system, say Eqs.~(\ref{eq:eom}), is defined as \beq \D=f-\sum_k\left(
t_k-l_k+r_k\right)=f-t+l-r,\eeq{compatibility} where $t_k,\,l_k$ and $r_k$ are respectively the number of equations,
independent gauge identities and gauge symmetries of order $k$. If $\D=0$, the system is said to be absolutely compatible.

Again, the Proca system~(\ref{eq:involutiveeom}) in $d$ dimension consists of $d$ second order equations $T^\m$, and another
first order one $T$. As we have seen, this system has a gauge identity~(\ref{gaugeid01}), but no gauge symmetries. Because the
field $\vf_\mu$ contains $f=d$ components to begin with, one finds from definition~(\ref{compatibility}) that $\D=d-(d+1)+1=0$,
so that the system~(\ref{eq:involutiveeom}) is absolutely compatible. The same is true for arbitrary spin, as we will show in
Section~\ref{sec:EMs}.

In fact, all known physical systems are absolutely compatible. It is plausible that any reasonable field theory has $\D=0$.
If an involutive system of equations is absolutely compatible, the number of DoFs it describes is given by~\cite{KaLySh3}:
\beq \mathfrak D=\tfrac{1}{2}\sum_k k\left(t_k-l_k-r_k\right),\eeq{eq:dof} provided both the gauge symmetry and gauge identity
generators are irreducible. This simple formula enables one to covariantly control the number of physical DoFs.

Accordingly, the involutive Proca system~(\ref{eq:involutiveeom}) has the DoF count
\beq \mathfrak D=\tfrac{1}{2}\left[1\times(1-0-0)+2\times(d-0-0)+3\times(0-1-0)\right]=d-1,\nonumber\eeq{xxxxxx}
which is indeed the correct number of physical polarizations of a massive spin-1 field. For massive particles of arbitrary
spin, we will see in Section~\ref{sec:EMs} that the counting~(\ref{eq:dof}) matches with the formula~(\ref{int4}),
as expected.

\subsubsection*{Consistent Deformations}

As suggested in Ref.~\cite{KaLySh3}, one can control the consistency of interactions by exploiting the involutive form and
the gauge symmetries and identities of the free field equations. Given an original system of free fields, the following
procedure enables one to introduce consistent couplings at the level of EoMs.
\begin{enumerate}

\item The free system of equations are written down in an involutive form.

\item All the gauge symmetries and identities of the free involutive system are identified.

\item Interactions are realized through deformations of the equations, gauge symmetries and identities.
Perturbatively in some coupling constant $\l$, the deformations are:
\bea T^a&=&T_0^a+\l T_1^a+\l^2 T_2^a+\cdots,\nonumber\\
L^A_a&=&L^A_{0a}+\l L^A_{1a}+\l^2 L^A_{2a}+\cdots,\label{def2}\\
R^i_\a&=&R^i_{0\a}+\l R^i_{1\a}+\l^2 R^i_{2\a}+\cdots.\nonumber\eea{def3}

\item The deformations~(\ref{def2}) are chosen such that at every order in $\l$ three requirements are fulfilled:
(a) the system remains involutive and absolutely compatible;
(b) the deformed system has the same number of gauge symmetry and identities, i.e., the quantities $l=\sum l_k$ and
$r=\sum r_k$ remain the same;
(c) the number of physical polarizations given by Eq.~(\ref{eq:dof}) remains the same as in the free theory.
\end{enumerate}

The requirements (a) and (b) guarantee that the system remains algebraically consistent with perturbatively included
interactions, while condition (c) ensures the deformed system has the same number of physical DoFs as the original
one\footnote{Notice that the orders of the equations, gauge identities and symmetries may increase (but never
decrease) at any order in $\l$. One should be careful about possible inclusion of higher derivative kinetic terms,
which may signal propagating ghosts.}.

\subsection*{Warm-Up Example: Spin 1}

The free involutive Proca system~(\ref{eq:involutiveeom}) and its gauge identity generators~(\ref{eq:gidgens1})
can be minimally coupled to EM by the substitution $\de_\m\rightarrow D_\m=\de_\m+ieA_\m$, so that we have
\bea &T_0^\m=\left(D^2-m^2\right)\vf^\m=0,\qquad T_0=D\cdot\vf=0,&\label{1spin1}\\
&L_0^\m=D^\m,\qquad L_0=-\left(D^2-m^2\right).&\eea{2spin1}
This is indeed the zeroth order deformation in the coupling constant $e$, because the associated gauge identity
fails or become anomalous only at $\mathcal O(e)$: \beq L_0\trr T_0\equiv L_0^\m T_{0\m}+L_0T_0=[D_\m,D^2]\vf^\m
=\mathcal{O}(e),\eeq{3spin1} due to the non-commutativity of the covariant derivatives.
This is precisely the anomaly noticed by Fierz and Pauli~\cite{pf}. But this failure can be rectified at $\mathcal O(e)$
by the inclusion of appropriate first order deformations, such that \beq \left(L_0+L_1\right)\trr\left(T_0+T_1\right)
=L_0\trr T_0+L_0\trr T_1+L_1\trr T_0+L_1\trr T_1=\mathcal O(e^2).\eeq{5spin1} That is, the first order deformations
must obey \beq L_0\trr T_1+L_1\trr T_0=-L_0\trr T_0+\mathcal O(e^2)=[D^2,D_\m]\vf^\m+\mathcal O(e^2).\eeq{6spin1}
One can explicitly compute the commutator on the right hand side; it is given by
\beq [D^2,D_\m]\vf^\m=D_\m\left(2ieF^{\m\n}\vf_\n\right)-ie\de_\m F^{\m\n}\vf_\n.\eeq{7spin1}
The left hand side of consistency condition~(\ref{6spin1}) can also be made more explicit: \beq L_0\trr T_1
+L_1\trr T_0=D_\m T_1^\m-\left(D^2-m^2\right)T_1+L_1^\m\left(D^2-m^2\right)\vf_\m+L_1(D\cdot\vf).\eeq{8spin1}
One can now compare Eqs.~(\ref{7spin1}) and~(\ref{8spin1}) to identify $T_1^\m=2ieF^{\m\n}\vf_\n$. The term
$\de_\m F^{\m\n}\vf_\n$, however, cannot be identified with anything else if one wants to avoid non-local
deformations containing the operator $\left(D^2-m^2\right)^{-1}$. Local deformations are still possible if
the photon is a background that obeys the source-free Maxwell equations: \beq \de_\m F^{\m\n}=0.\eeq{Maxwell}
Then, a set of consistent deformations up to $\mathcal O(e)$ is given by
\bea &T^\m=\left(D^2-m^2\right)\vf^\m+2ieF^{\m\n}\vf_\n=0,\qquad T=D\cdot\vf=0,&\label{9spin1}\\
&L^\m=D^\m,\qquad L=-\left(D^2-m^2\right).&\eea{10spin1}
Actually, these deformations are correct up to all orders since $L\trr T$ vanishes. It is easy to see that the
deformed system~(\ref{9spin1}) is involutive. Actually, the individual values of $t_k$ and $l_k$ do not change
$\forall k$. Therefore, this system is algebraically consistent and describes the same number of DoFs
as the free theory, namely $d-1$. The propagation of these DoFs is manifestly causal since the deformed equations
contain no higher derivative kinetic terms.

Similarly, for gravitational coupling the zeroth order deformations\footnote{In this case, one may think that the
Riemann curvature incorporates the deformation parameter.} are obtained from Eqs.~(\ref{eq:involutiveeom})
and~(\ref{eq:gidgens1}) by the minimal substitution $\de_\m\rightarrow\nb_\m$. That is,
\bea &T_0^\m=\left(\nb^2-m^2\right)\vf^\m=0,\qquad T_0=\nb\cdot\vf=0,&\label{1gspin1}\\
&L_0^\m=\nb^\m,\qquad L_0=-\left(\nb^2-m^2\right),&\eea{2gspin1} where $[\nb_\m,\nb_\n]\vf^\r=R^\r_{~\s\m\n}\vf^\s$.
The gauge identity anomaly in this case is given by \beq  L_0\trr T_0=-[\nb^2,\nb_\m]\vf^\m=\nb_\m\left(R^{\m\n}
\vf_\n\right).\eeq{3gspin1} This anomaly is cured up to all orders, without any restrictions on the gravitational
field, by first order deformations with only $T_1^\m=-R^{\m\n}\vf_\n$ non-vanishing:
\bea &T^\m=\left(\nb^2-m^2\right)\vf^\m-R^{\m\n}\vf_\n=0,\qquad T=\nb\cdot\vf=0,&\label{9gspin1}\\
&L^\m=\nb^\m,\qquad L=-\left(\nb^2-m^2\right).&\eea{10gspin1}
The Proca field therefore interacts consistently with an arbitrary gravitational field because indeed the above
deformations identically satisfy the gauge identity to all orders:
\beq L\trr T=-[\nb^2,\nb_\m]\vf^\m-\nb_\m\left(R^{\m\n}\vf_\n\right)=0.\eeq{11gspin1}

Having worked out the simple but instructive examples of spin 1, we are now ready to consider the EM and gravitational
couplings of arbitrary-spin particles.

\section{Arbitrary Spin: Non-Minimal Couplings}\label{sec:EMs}

In this Section we are going to deform the free massive HS system to construct consistent couplings to EM and gravitational
backgrounds, using the formalism previously explained. As we go along, we will develop a methodology meant for deformations
of this particular system. The same methodology works for EM as well as for gravitational couplings.

\subsection*{Free Involutive System}\label{sec:Frees}

The staring point is the free involutive system for a massive spin-$s$ particle, which is
\bea &T_{\m_1...\m_s}=(\de^2-m^2)\vf_{\m_1...\m_s}=0,&\label{free1a}\\
&T_{\m_1...\m_{s-1}}=\de\cdot\vf_{\m_1...\m_{s-1}}=0.&\eea{free1b}
Note that, unlike the Fierz-Pauli conditions~(\ref{int1})-(\ref{int3}), this system does not incorporate the trace
constraint as a zeroth order differential equation. In fact, the field $\vf_{\m_1...\m_s}$ appearing in the involutive
system~(\ref{free1a})-(\ref{free1b}) is an irreducible representation of the Lorentz group: a symmetric \emph{traceless}
rank-$s$ tensor. Because of this reason, the number of second order equations in the system~(\ref{free1a})-(\ref{free1b})
is the same as the number of independent components of a rank-$s$ symmetric traceless tensor\footnote{This is also the
number of fields $f$ described by the system.}, which is \beq t_2={d-1+s\choose s}-{d-3+s\choose s-2}.\eeq{free2a}
Similarly, the transversality condition amounts to \beq t_1={d-2+s\choose s-1}-{d-4+s\choose s-3}\eeq{free2b}
first order equations. On the other hand, the system possesses third order gauge identities:
\beq \de^{\m_s}T_{\m_1...\m_s}-(\de^2-m^2)T_{\m_1...\m_{s-1}}=[\de^{\m_s},\de^2-m^2]\,\vf_{\m_1...\m_s}=0.\eeq{free04}
In the compact form~(\ref{eq:lgen}), they read
\beq L^{\a_1\dots\a_{s-1}}\trr T=[\de_\m,\de^2]\,\vf^{\m\a_1...\a_{s-1}}=0,\eeq{free4}
where the gauge identity generators are given by
\bea &L^{\a_1\dots\a_{s-1},\m_1\dots\m_s}=\eta^{\a_1\dots\a_{s-1},(\m_1\dots\m_{s-1}}\de^{\m_s)},&\label{free3a}\\
&L^{\a_1\dots\a_{s-1},\m_1\dots\m_{s-1}}=-\eta^{\a_1\dots\a_{s-1},\m_1\dots\m_{s-1}}(\de^2-m^2).&\eea{free3b}
Notice that the trace of the identity~(\ref{free4}) is vanishing on account of the tracelessness of the field itself.
The number of independent gauge identities is again given by that of the independent components of a of a rank-$(s-1)$
symmetric traceless tensor, namely \beq l_3={d-2+s\choose s-1}-{d-4+s\choose s-3}.\eeq{free5}
Finally, the system~(\ref{free1a})-(\ref{free1b}) enjoys no gauge symmetries whatsoever. To summarize, at $k$-th
order in derivatives, the number of equations $t_k$, independent gauge identities $l_k$, and gauge symmetries $r_k$
are respectively given by
\beq t_k=t_1\d_k^1+t_2\d_k^2,\qquad l_k=l_3\d_k^3,\qquad r_k=0.\eeq{free6} Clearly, the compatibility
coefficient~(\ref{compatibility}) vanishes, \beq \Delta=f-t_1-t_2+l_3=0,\eeq{free7} on account of the equalities
$f=t_2$ and $t_1=l_3$. That is, the system is absolutely compatible. Then the number of physical DoFs is given by
formula~(\ref{eq:dof}); it is \beq \mathfrak D=\tfrac{1}{2}\sum_k k(t_k-l_k)=\tfrac{1}{2}\left(t_1+2t_2-3l_3\right)
=t_2-t_1.\eeq{free8} In view of the expressions~(\ref{free2a}),~(\ref{free2b}) and~(\ref{free5}), this coincides
with the formula~(\ref{int4}) for the propagating DoFs of a massive spin-$s$ particle in $d$ dimensions.

\subsection*{Deformation in EM Background}\label{sec:Defms}

As we have identified the gauge identities of the free system, we would like to exploit its involutive form to
introduce consistent interactions. First, we consider EM coupling. With this end in view, we consider the
deformations of the equations,
\bea T^{\m_1\dots\m_s}_\text{free}\rightarrow T^{\m_1\dots\m_s}&=&\sum_{n=0}^\infty T_n^{\m_1\dots\m_s}=0,
\label{defm1}\\T^{\m_1\dots\m_{s-1}}_\text{free} \rightarrow T^{\m_1\dots\m_{s-1}}&=&\sum_{n=0}^\infty
T_n^{\m_1\dots\m_{s-1}}=0,\eea{defm2}
and also of the gauge identity generators,
\bea L_\text{free}^{\a_1\dots\a_{s-1},}{}_{\m_1\dots\m_s}\rightarrow L^{\a_1\dots\a_{s-1},}{}_{\m_1\dots\m_s}&=&
\sum_{n=0}^\infty L_n^{\a_1\dots\a_{s-1},}{}_{\m_1\dots\m_s},\label{defm3}\\L_\text{free}^{\a_1\dots\a_{s-1},}
{}_{\m_1\dots\m_{s-1}}\rightarrow L^{\a_1\dots\a_{s-1},}{}_{\m_1\dots\m_{s-1}}&=&\sum_{n=0}^\infty L_n^{\a_1
\dots\a_{s-1},}{}_{\m_1\dots\m_{s-1}},\eea{defm4}
where $n$ denotes the order of the perturbative expansion in the EM charge $e$. We require that the deformed system
satisfy the gauge identities
\beq L^{\a_1\dots\a_{s-1}}\trr T=\sum_{m=0}^\infty\sum_{n=0}^\infty\left(L_m^{\a_1\dots\a_{s-1}}\trr T_n\right)=0.
\eeq{defm5}
If the system~(\ref{defm1})-(\ref{defm2}) remains involutive, then at each order of the deformation the equations (i.e.,
$\forall n$ the quantities $T_n^{\m_1\dots\m_s}$ and $T_n^{\m_1\dots\m_{s-1}}$) must be symmetric and traceless\footnote{
Otherwise, at any given order one would find unwarranted constraints on the field, which vanish when the interaction is
turned off. The involutive form eliminates this unacceptable possibility.}. Similarly, the gauge identities~(\ref{defm5})
must also remain symmetric and traceless since otherwise it would mean an unwarranted change in their total number.

The deformed gauge identities~(\ref{defm5}) break down into a cascade, order by order in $e$. The zeroth order gauge
identities may become anomalous at $\mathcal O(e)$, and they read
\beq \mathcal A^{\a_1\dots\a_{s-1}}\equiv-L_0^{\a_1\dots\a_{s-1}}\trr T_0=\mathcal O(e),\eeq{anomaly}
where $\mathcal A^{\a_1\dots\a_{s-1}}$ is called the (symmetric traceless) anomaly tensor. At first order this
anomaly must be rectified so as to push the failure of the gauge identities to $\mathcal{O}(e^2)$. The first order
gauge identities can be rewritten as
\beq \mathcal A^{\a_1\dots\a_{s-1}}=L_0^{\a_1\dots\a_{s-1}}\trr T_1+L_1^{\a_1\dots\a_{s-1}}\trr T_0+\cdots\,,\eeq{boss}
where the ellipses stand for $\mathcal O(e^2)$ terms. It is important to note that if there are no such $\mathcal O(e^2)$
terms, then the system requires no higher order deformations provided that
\beq \mathcal C^{\a_1\dots\a_{s-1}}\equiv L_1^{\a_1\dots\a_{s-1}}\trr T_1=0.\eeq{last}

Eqs.~(\ref{anomaly})-(\ref{last}) constitute the core of our analysis. The program is to compute the anomaly
tensor~(\ref{anomaly}) and rewrite it in a suitable form, so that one can read off the first order deformations
from Eq.~(\ref{boss}). If these identifications leave us with no second order terms and if Eq.~(\ref{last}) is
satisfied, then the deformations consistently stop at first order.

Note that the zeroth order equations are given not by the free system~(\ref{free1a})-(\ref{free1b}) itself, but by
the minimally coupled version resulting from $\de_\m\rightarrow D_\m$ for EM interactions:
\bea &T_0^{\m_1...\m_s}=(D^2-m^2)\vf^{\m_1...\m_s}=0,&\label{EM1a}\\
&T_0^{\m_1...\m_{s-1}}=D\cdot\vf^{\m_1...\m_{s-1}}=0,&\eea{EM1b}
while the zeroth order gauge identity generators follow similarly from Eqs.~(\ref{free3a})-(\ref{free3b}):
\bea L_0^{\a_1\dots\a_{s-1},\m_1\dots\m_s}&=&\eta^{\a_1\dots\a_{s-1},(\m_1\dots\m_{s-1}}D^{\m_s)},\label{eq:Larbspin1}\\
L_0^{\a_1\dots\a_{s-1},\m_1\dots\m_{s-1}}&=&-\eta^{\a_1\dots\a_{s-1},\m_1\dots\m_{s-1}}(D^2-m^2).\eea{eq:Larbspin2}
These indeed qualify as the correct zeroth order deformations since the anomaly tensor is
\beq \mathcal A^{\a_1\dots\a_{s-1}}=[D^2,D_\m]\,\vf^{\m\a_1\dots\a_{s-1}}=\mathcal{O}(e).\eeq{anoalyEM}

Next, one has to rewrite this expression for the anomaly tensor in a form that facilitates the comparison with the first
order deformations through Eq.~(\ref{boss}). We are particularly interested in finding the non-minimal couplings that
show up as corrections to the Klein-Gordon equation~(\ref{EM1a}). In Eq.~(\ref{boss}) they appear in the following form
\beq L_0^{\a_1\dots\a_{s-1},}{}_{\m_1\dots\m_s}T_1^{\m_1...\m_s}=D_{\a_s}T_1^{\a_1\dots\a_s}.\eeq{non-minimal}
Therefore, we should extract from the anomaly tensor total derivatives of symmetric traceless objects. Divergence of
the field may also appear in Eq.~(\ref{boss}) through the term
\beq L_1^{\a_1\dots\a_{s-1},}{}_{\m_1\dots\m_{s-1}}T_0^{\m_1...\m_{s-1}}=L_1^{\a_1\dots\a_{s-1},}{}_{\m_1\dots\m_{s-1}}
D\cdot\vf^{\m_1...\m_{s-1}}.\eeq{divergence} The anomaly tensor may also give rise to terms which are neither of the
above two forms. The sum $\mathcal B^{\a_1\dots\a_{s-1}}$  of all such terms must be identified through Eq.~(\ref{boss})
as \beq \mathcal B^{\a_1\dots\a_{s-1}}=L_1^{\a_1\dots\a_{s-1},}{}_{\m_1\dots\m_s}(D^2-m^2)\,\vf^{\m_1...\m_s}
-(D^2-m^2)\,T_1^{\a_1\dots\a_{s-1}}.\eeq{lincomb} This identification, however, can only give non-local solutions for
the relevant first order deformations\footnote{We will comment on non-locality in the next Section.}. Problematic for
local deformations, such terms must vanish if locality has to be preserved in the absence of additional DoFs. This
may impose some restrictions on the external background. Constraints on the field, however, must be avoided lest
the very involutive form of the system should be ruined.

In order to rewrite the anomaly tensor~(\ref{anoalyEM}), we use properties of the commutator and the product rule
for covariant derivatives. Thus we obtain \beq \mathcal A^{\a_1\dots\a_{s-1}}=-2ie D_\m\left(F^{\r\m}\vf_\r^{~\a_1
\dots\a_{s-1}}\right)-ie\de_\m F^{\m\n}\vf_\n^{~\a_1\dots\a_{s-1}}.\eeq{an1} The total derivative term can be cast
into the form~(\ref{non-minimal}) if the rank-$s$ tensor inside the parentheses is made symmetric and traceless.
While this is done at the cost of adding and subtracting some total derivatives, the left-over terms give rise to
divergence pieces of the form~(\ref{divergence}) on account of the product rule. The final result is
\bea \mathcal A^{\a_1\dots\a_{s-1}}&=& D_\m\left[-2iesF^{\r(\m}\vf_\r^{~\a_1\dots\a_{s-1})}\right] +2ie(s-1)F^{\r(\a_1}
D\cdot\vf_\r^{~\a_2\dots\a_{s-1})}\nonumber\\&&+2ie(s-1)\de_\m F_\n^{~(\a_1}\vf^{\a_2\dots\a_{s-1})\m\n}-ie\de_\m F^{\m\n}
\vf_\n^{~\a_1\dots\a_{s-1}}.\eea{an2}
In view of Eq.~(\ref{boss}), the first line of the above expression gives us the identifications
\bea &T_1^{\m_1\dots\m_s}=-2ies F^{\r(\m_1}\vf_\r^{~\m_2\dots\m_s)},&\label{T1EM}\\&L_1^{\a_1\dots\a_{s-1},}
{}_{\m_1\dots\m_{s-1}}=-2ie(s-1)\,\d^{\a_1\dots\a_{s-1}}_{\r(\m_1\dots\m_{s-2}}F^\r_{~\m_{s-1})},&\eea{L1EM}
thanks to Eqs.~(\ref{non-minimal}) and~(\ref{divergence}).
On the other hand, the second line of Eq.~(\ref{an2}) is identified as $\mathcal B^{\a_1\dots\a_{s-1}}$, which
should vanish, as we already pointed out from Eq.~(\ref{lincomb}). Splitting the gradient of the field strength into
irreducible Lorentz tensors, one obtains
\beq \mathcal B^{\a_1\dots\a_{s-1}}=2ie(s-1)Q_{\m\n}{}^{(\a_1}\vf^{\a_2\dots\a_{s-1})\m\n}-ie\left(\tfrac{2s+d-3}{d-1}
\right)\de_\m F^{\m\n}\vf_\n^{~\a_1\dots\a_{s-1}},\eeq{badEM}
where $Q_{\m\n}{}^\a$ is the symmetric traceless gradient of the field strength in $d$ dimensions:
\beq Q_{\m\n}{}^\a\equiv\de_{(\m}F_{\n)}{}^\a-\left(\tfrac{1}{d-1}\right)\left[\h_{\m\n}\de_\r F^{\r\a}+
\d^\a_{(\m}\de^\r F_{\n)\r}\right],\eeq{badEM1}
and the (anti)symmetric tensors $\de_{(\m}F_{\n\a)}$ and $\de_{[\m}F_{\n\a]}$ are zero identically. For $s>1$,
it is clear from Eq.~(\ref{badEM}) that $\mathcal B^{\a_1\dots\a_{s-1}}$ can be zero, without incurring
unwarranted constraints on the HS field, if and only if both the quantities $Q_{\m\n}{}^\a$ and $\de_\m F^{\m\n}$ vanish.
The above conditions are tantamount to requiring that the EM background satisfy: \beq \de_{(\m}F_{\n)\r}=0.\eeq{bckEM}
This admits, in particular, $F_{\m\n}=\text{constant}$ as a consistent background.\footnote{Non-constant EM backgrounds
may also qualify. In $d=4$, for example, the generic solution of Eq.~(\ref{bckEM}) is: $F^{0i}=\e^i+\varepsilon^{ijk}\a_j x_k$
and $F^{ij}=\varepsilon^{ijk}\left(\b_k+\a_k t\right)$, with $\e_i, \a_i, \b_i$ constants and $i=1,2,3$.}
For $s=1$, however, it suffices to require that the background obey the source-free Maxwell equations: $\de_\m F^{\m\n}=0$.
Finally, the deformations can be made consistent up to all orders by the choice:
\beq T_1^{\m_1\dots\m_{s-1}}=0,\qquad L_1^{\a_1\dots\a_{s-1},}{}_{\m_1\dots\m_s}=0.\eeq{choiceEM} Indeed, this choice renders
the tensor $\mathcal C^{\a_1...\a_{s-1}}$ appearing in Eq.~(\ref{last}) vanishing.

To summarize, we have found the following consistently deformed involutive system:
\bea &T^{\m_1...\m_s}=(D^2-m^2)\vf^{\m_1...\m_s}-2ies F^{\r(\m_1}\vf_\r^{~\m_2\dots\m_s)}=0,&\label{EMfinala}\\[.2cm]
&T^{\m_1...\m_{s-1}}=D\cdot\vf^{\m_1...\m_{s-1}}=0,&\eea{EMfinalb}
for a class of EM backgrounds. Augmented by the implicit trace condition, $\vf'_{\m_1...\m_{s-2}}=0$, the same equations
show up, quite curiously, in string theory as well~\cite{AN2,PRS}. This system is algebraically consistent by
construction. The DoF count is also correct for an obvious reason: this system and the free one shares the same set of
$t_k$ and $l_k$ given in Eq.~(\ref{free6}). The Laplacian kinetic operator in Eq.~(\ref{EMfinala}) also makes
causal propagation manifest.

\subsection*{Deformation in Gravitational Background}\label{sec:DefmsGR}

Now we turn to gravitational coupling. In this case, the Riemann curvature is assumed to have incorporated the
deformation parameter, and the covariant derivatives are denoted by $\nb_\m$. Modulo these, the steps and analyses 
of the previous Subsection hold verbatim in this case until one writes down an explicit expression like~(\ref{an1}) 
for the anomaly tensor. While \beq \mathcal A^{\a_1\dots\a_{s-1}}=[\nb^2,\nb_\m]\,\vf^{\m\a_1\dots\a_{s-1}},\eeq{GR1}
similar steps lead to the gravitational counterpart of Eq.~(\ref{an1}), which reads
\bea \mathcal A^{\a_1\dots\a_{s-1}}&=&\nabla_\m\left(2\sum_{i=1}^{s-1}R^\m{}_\n{}^{\a_i}{}_\r\,\vf^{\a_1\dots\a_{i-1}
\a_{i+1}\dots\a_{s-1}\n\r}-R^{\r\m}\vf_\r{}^{\a_1\dots\a_{s-1}}\right)\nonumber\\&&-\sum_{i=1}^{s-1}\nb_\m R^\m{}_\n{}^
{\a_i}{}_\r\,\vf^{\a_1\dots\a_{i-1}\a_{i+1}\dots\a_{s-1}\n\r}.\eea{GR2} In deriving the above we have used the symmetry
properties of the Riemann tensor, which imply in particular $R^\m{}_{(\r}{}^\n{}_{\s)}=R^{(\m}{}_\r{}^{\n)}{}_\s$.
Again, from the total derivative term in the first line of Eq.~(\ref{GR2}) one can extract the gravitational
counterpart of Eq.~(\ref{non-minimal}), namely
\beq L_0^{\a_1\dots\a_{s-1},}{}_{\m_1\dots\m_s}T_1^{\m_1...\m_s}=\nb_{\a_s}T_1^{\a_1\dots\a_s}.\eeq{non-minimalGR}
The feat is achieved by rendering the rank-$s$ tensor
inside the parentheses symmetric and traceless. The left-over terms from the last step again produce covariant
divergences$-$the gravitational counterpart of Eq.~(\ref{divergence})$-$of the form
\beq L_1^{\a_1\dots\a_{s-1},}{}_{\m_1\dots\m_{s-1}}T_0^{\m_1...\m_{s-1}}=L_1^{\a_1\dots\a_{s-1},}{}_{\m_1\dots\m_{s-1}}
\nb\cdot\vf^{\m_1...\m_{s-1}},\eeq{divergenceGR} thanks to the product rule. On the other hand, one can massage
the second line of Eq.~(\ref{GR2}) by using the contracted Bianchi identity:
$\nb_\m R^\m{}_\n{}^\a{}_\r=\nb^\a R_{\n\r}-\nb_\r R_\n^{~\a}$.
All these steps lead us to the following expression for the anomaly tensor:
\bea \mathcal A^{\a_1\dots\a_{s-1}}&=&\nb_\m\left[s(s-1)R^{(\m}{}_{\n}{}^{\a_1}{}_{\r}\vf^{\a_2\dots\a_{s-1})\n\r}
-sR^{\r(\m}\vf_{\r}{}^{\a_1\dots\a_{s-1})}\right]\nonumber\\&&-(s-1)\left[(s-2)R^{(\a_1}{}_{\m}{}^{\a_2}{}_{\n}\nb\cdot
\vf^{\a_3\dots\a_{s-1})\m\n}-R^{\m(\a_1}\nb\cdot\vf_{\m}{}^{\a_2\dots\a_{s-1})}\right]\nonumber\\&&+(s-1)\left[2\nb_\m
R_\n^{~(\a_1}\vf^{\a_2\dots\a_{s-1})\m\n}-\nb^{(\a_1}R_{\m\n}\vf^{\a_2\dots\a_{s-1})\m\n}\right]\nonumber\\
&&-(s-1)(s-2)\nb_\m R^{(\a_1}{}_{\n}{}^{\a_2}{}_{\r}\,\vf^{\a_3\dots\a_{s-1})\m\n\r}.\eea{GR3}
When plugged into the first order gauge identity~(\ref{boss}), the first line of this expression gives us, in view
of Eq.~(\ref{non-minimalGR}), the following identification for the deformation of equations:
\beq T_1^{\m_1\dots\m_s}=s(s-1)R^{(\m_1}{}_{\n}{}^{\m_2}{}_{\r}\vf^{\m_3\dots\m_s)\n\r}-sR^{\r(\m_1}\vf_{\r}{}^{\m_2
\dots\m_s)}.\eeq{T1GR}
Similarly, from the second line one identifies, on account of Eq.~(\ref{divergenceGR}),
\beq L_1^{\a_1\dots\a_{s-1},}{}_{\m_1\dots\m_{s-1}}=-(s-1)\left[(s-2)\,\d^{\a_1\dots\a_{s-1}}
_{\r\s(\m_1\dots\m_{s-3}}R^\r{}_{\m_{s-2}}{}^\s{}_{\m_{s-1})}-\d^{\a_1\dots\a_{s-1}}_{\r(\m_1\dots\m_{s-2}}
R^\r{}_{\m_{s-1})}\right].\eeq{GR4}

The remaining third and fourth lines in the expression~(\ref{GR3}) for the anomaly tensor are identified as
$\mathcal B^{\a_1\dots\a_{s-1}}$, which must be set to zero in order to avoid non-locality. Now
$\mathcal B^{\a_1\dots\a_{s-1}}$ contains gradients of the Riemann and Ricci tensors, and the latter quantities
can be split into irreducible Lorentz tensors, so that one obtains the expression
\bea \mathcal B^{\a_1\dots\a_{s-1}}&=&-\tfrac{(s-1)(s-2)}{d-2}\left[(d-2)X_{\m\n\r}{}^{(\a_1\a_2}\,\vf^{\a_3\dots
\a_{s-1})\m\n\r}+Y_{\m\n\r}\,g^{(\a_1\a_2}\vf^{\a_3\dots\a_{s-1})\m\n\r}\right]\nonumber\\[4pt]&&+\left(\tfrac{s-1}
{d-2}\right)\left[(2s+d-6)\,Y_{\m\n}{}^{(\a_1}\vf^{\a_2\dots\a_{s-1})\m\n}-\left(\tfrac{s+2d-6}{3}\right)
Z_{\m\n}{}^{(\a_1}\vf^{\a_2\dots\a_{s-1})\m\n}\right]\nonumber\\[5pt]&&+\tfrac{2(s-1)(s+d-2)}{(d-1)(d+2)}
\left(\nb_\m R\right)\vf^{\a_1\dots\a_{s-1}\,\m},\eea{badGR}
where $X_{\m\n\r}{}^{\a\b}$, $Y_{\m\n}{}^\a$ and $Z_{\m\n}{}^\a$ are the following irreducible Lorentz tensors:
\bea X_{\m\n\r}{}^{\a\b}&=&\nb_{(\m}W_\n{}^\a{}_{\r)}{}^\b-\left(\tfrac{2}{d+2}\right)g_{(\m\n}\nb^\s W_{\r)}{}^{(\a}
{}_\s{}^{\b)},\label{Xdefined}\\Y_{\m\n\r}&=&\nb_{(\m}R_{\n\r)}-\left(\tfrac{2}{d+2}\right)g_{(\m\n}\nb_{\r)}R,
\label{Ydefined}\\Z_{\m\n\r}&=&2\nb_{[\r}R_{\m]\n}+\left(\tfrac{1}{d-1}\right)g_{\n[\r}\nb_{\m]}R+\left(\m
\leftrightarrow\n\right),\eea{Zdefined}
with $W_{\m\a\n\b}$ denoting the Weyl tensor.\footnote{The quantities ~(\ref{Xdefined})-(\ref{Zdefined}) are all
traceless, thanks to the identity $\nb^\m R_{\m\n}=\tfrac{1}{2}\nb_\n R$.} Now that $\mathcal B^{\a_1\dots\a_{s-1}}$
must be set to zero without imposing any further constraints on the HS field, we have three different cases:
\begin{itemize}
\item $s=1$: Because all the dangerous terms in Eq.~(\ref{badGR}) are proportional to $s-1$, they vanish automatically
for the Proca field and pose no restrictions on the background.

\item $s=2$: In this case, the first line in Eq.~(\ref{badGR}) vanishes. In order to kill the other terms, one must
set to zero all the quantities $Y_{\m\n\r}$, $Z_{\m\n\r}$ and $\nb_\m R$, which is tantamount to having a
covariantly constant Ricci tensor:
\beq \nb_\m R_{\n\r}=0.\eeq{spin2-good}
Thus, consistency requires that the gravitational background be Ricci symmetric.

\item $s\geq3$: For higher spins, on top of having a Ricci symmetric space, one needs additional conditions on the Weyl
tensor, namely $X_{\m\n\r}{}^{\a\b}=0$. Because the divergence of the Weyl tensor can be expressed in terms of $Z_{\m\n\r}$
and $\nb_\m R$ as a consequence of the Bianchi identities, we have the equivalent set of conditions:
\beq \nb_\m R_{\n\r}=0,\qquad \nb_{(\m}W_\n{}^\a{}_{\r)}{}^\b=0,\eeq{spins-good}
which a gravitational background must satisfy in order to propagate consistently an arbitrary-spin particle in isolation,
under the assumption of locality. Note, in particular, that symmetric spaces do qualify as consistent backgrounds, since
they have covariantly constant Riemann tensors: $\nb_\m R_{\n\a\r\b}=0$.
\end{itemize}

Given one of the appropriate restrictions, one can render the deformations~(\ref{T1GR}) and~(\ref{GR4}) consistent up
to all orders by simply choosing
\beq T_1^{\m_1\dots\m_{s-1}}=0,\qquad L_1^{\a_1\dots\a_{s-1},}{}_{\m_1\dots\m_s}=0.\eeq{choiceGR} Surely, the tensor
$\mathcal C^{\a_1...\a_{s-1}}$ appearing in Eq.~(\ref{last}) vanishes with this choice.
Thus, we end up having the following consistently deformed involutive system:
\bea &T^{\m_1...\m_s}=(\nb^2-m^2)\vf^{\m_1...\m_s}+\left[s(s-1)R^{(\m_1}{}_{\n}{}^{\m_2}{}_{\r}\vf^{\m_3\dots\m_s)\n\r}
-sR^{\r(\m_1}\vf_{\r}{}^{\m_2\dots\m_s)}\right]=0,~~~~~~&\label{GRfinala}\\[.25cm]
&T^{\m_1...\m_{s-1}}=\nb\cdot\vf^{\m_1...\m_{s-1}}=0,&\eea{GRfinalb}
with the aforementioned restrictions on the gravitational background. The algebraic consistency, preservation of the correct
number of DoFs and causal propagation are guaranteed precisely the same way as they are in an EM background and in the free theory.

\section{On $g$- \& $h$-Factors and Non-Locality}\label{sec:ghNL}

The EM and gravitational non-minimal couplings we found in Section~\ref{sec:EMs} are respectively called the magnetic
dipole and the gravitational quadrupole terms. The magnetic dipole moment is quantified by the so-called gyromagnetic
ratio or the $g$-factor. For a spin-$s$ boson of mass $m$ and charge $e$, the $g$-factor appears at the level of EoM
as follows~\cite{d3}: \beq (D^2-m^2)\vf^{\m_1...\m_s}-iegs F^{\r(\m_1}\vf_\r^{~\m_2\dots\m_s)}=0.\eeq{g-factor}
Direct comparison of this with our result~(\ref{EMfinala}) reveals that, for all spins, \beq g=2.\eeq{gis2}

This may not come as a surprise, since $g=2$ turns out to be the ``preferred'' tree-level value for any
spin~\cite{Weinberg,g=2}. Moreover, open string theory predicts the same universal value for $g$~\cite{AN2,PRS,g=2}.
On the other hand, it has been observed that Kaluza-Klein (KK) reductions of consistent higher dimensional
models give $g=1$ for all spins~\cite{KK1,KK2,KK3}. How does this fact go along with our results?

The answer lies in non-locality$-$a possibility we did not explore. In fact, KK theories describe a tower of massive
particles, not a single one. If one is interested in the dynamics of a particular state, one may integrate out the
other ones, some of which are of comparable mass. This results in a non-local theory. The conclusion is that additional
dynamical states and/or non-local terms may change the value of $g$. Indeed, sum rules from low-energy Compton scattering
can show that in the presence of other massive state $g-2$ may become $\mathcal O(1)$~\cite{g_sumrule}. Let us see how
this could be understood within our framework.

For simplicity, let us consider $s=2$. Under the condition~(\ref{bckEM}), the anomaly tensor~(\ref{an2}) can be
rewritten, for an arbitrary gyromagnetic ratio $g$, as
\beq \mathcal A^\a= D_\m\left[-2iegF^{\r(\m}\vf_\r^{~\a)}\right]+iegF^{\r\a}D\cdot\vf_\r-ie(g-2)F^{\m\n}D_\m\vf_\n^\a,
\eeq{spin2-2}
The first two terms on the right hand side can again be incorporated into local first order deformations. Note,
in particular, that the first term gives rise to the value $g$ for the gyromagnetic ratio. Even when the background
satisfies the condition~(\ref{bckEM}), the last term with $g\neq 2$ signals breakdown of locality. More explicitly,
the consistency of the
first order gauge identity~(\ref{boss}) requires the identification \beq ie(g-2)F^{\m\n}D_\m\vf_\n^\a=-L_0^{\a,}{}_\m
T_1^\m-L_1^{\a,}{}_{\m\n} T_0^{\m\n}=(D^2-m^2)T_1^\a-L_1^{\a,}{}_{\m\n}(D^2-m^2)\vf^{\m\n}.\eeq{spin2-3} This may
admit only non-local solutions for the relevant first order deformations, like
\beq L_1^{\a,\m\n}=0,\qquad T_1^\m=ie(g-2)\left(\frac{1}{D^2-m^2}\right)\left(F^{\r\s}D_\r\vf_\s^\m\right),\eeq{spin2-4}
that modify the transversality condition.
The presence of the operator $(D^2-m^2)^{-1}$ is tantamount to non-locality, which might have arisen from integrating
out other massive states of the theory. Thus non-locality and/or additional DoFs of comparable mass may give $g\neq2$.
Non-local interactions, however, are beyond the scope of our present analysis.

On the other hand, the gravitational quadrupole moment is quantified analogously by the gravimagnetic ratio or the
$h$-factor. A careful definition of the $h$-factor was given in Ref.~\cite{GiLiPo}. At the level of EoM it shows up as
\beq (\nb^2-m^2)\vf^{\a(s)}+h\left[R_{\m\n\r\s}\,\tfrac{1}{2}{{(\S^{\m\n})}^{\a(s)}}_{\b(s)}\,\tfrac{1}{2}{{(\S^{\m\n})}
^{\b(s)}}_{\g(s)}\right]\vf^{\g(s)}+\dots=0,\eeq{h-factor0}
where the ellipses denote possible on-shell vanishing terms, and
\beq {(\S^{\m\n})_{\a(s)}}^{\b(s)}\equiv 2s\delta^{[\m}_{(\a_{1}}\eta^{\n](\b_{1}}\delta^{\b_{2}\dots\b_{s})}_{\a_{2}
\dots\a_{s})}=-{{(\S^{\m\n})}^{\b(s)}}_{\a(s)},\eeq{eq:sigmas}
are the components of the Lorentz generators in the spin-$s$ bosonic representation.

These are antisymmetric in $\m\n$, and symmetric in the indices of the individual sets $\a(s)$ and $\b(s)$.
These symmetry properties result in the tracelessness of $\S^{\m\n}$ in the $\a(s)$ indices and hence in the $\b(s)$ indices
as well\footnote{The definition of the Lorentz generators given in Ref.~\cite{GiLiPo}, though, is plagued with typographic
errors. We thank M.~Porrati for pointing this out and clarifying the properties of $\S^{\m\n}$. Note also that the
antisymmetry~(\ref{eq:sigmas}) between the $\a(s)$ and $\b(s)$ sets renders any cross contraction of indices vanishing.}.
A straightforward computation gives
\beq (\nb^{2}-m^{2})\vf^{\a_{1}\dots\a_{s}} +h\Bigl[s(s-1) R_{\m}{}^{(\a_1}{}_{\n}{}^{\a_2}\vf^{\a_3 \dots\dots\a_s)\m\n}
-sR^{\m(\a_1} \vf_{\m}{}^{\a_2\dots\a_s)}\Bigr]+\dots=0.\eeq{h-factor}
Upon comparing our results~(\ref{GRfinala}) with this definition, one immediately concludes \beq h=1.\eeq{h=1}

This also happens to be the preferred field theory value from considerations of tree-level unitarity and
supersymmetry~\cite{CPD}. However, the study of three-point functions in superstring theory suggests $h\neq1$ in
general~\cite{GiLiPo}, and this happens because of the existence of a whole tower of states of arbitrarily
large masses and spins. Quite similarly to the $g$-factor analysis, one can show for gravitational couplings
that $h\neq1$ necessarily results in non-locality, thanks to the consistency of the first order gauge
identity~(\ref{boss}). For a spin-2 particle in a Ricci symmetric space, the gravitational counterpart of the
identification~(\ref{spin2-3}) admits a non-local solution \beq L_1^{\a,\m\n}=0,\qquad T_1^\s=-(h-1)
\left(\frac{1}{\nb^2-m^2}\right)(2R^{\m\n\r\s}+R^{\m\n}g^{\r\s})\nb_\m\vf_{\n\r}.\eeq{spin2-10} Here the
presence of the operator $(\nb^2-m^2)^{-1}$ is a telltale sign of non-locality. This might be taken as a hint
for the existence of other interacting massive DoFs, which when taken into account, may restore locality in the
effective theory.

\section{Concluding Remarks}\label{sec:Conclusion}

In this paper, we have exploited properties of involutive differential equations to consistently couple an
arbitrary-spin massive particle to EM and gravitational backgrounds. Originally developed in~\cite{KaLySh3},
the method works at the level of field equations with a consistent perturbative scheme to introduce interactions.
The virtues of this framework are manifold$-$manifest covariance, built-in algebraic consistency, stability
of the number of physical DoFs and their manifest causal propagation, and striking simplicity.

It is surprising to see how easily this approach may produce non-trivial results and shed light on some
intricacies of interacting massive HS particles. Among others, we could reproduce the preferred field
theory values of the $g$- and $h$-factors, and see how these values may/do get altered in the
presence of other massive states. For EM coupling, we have seen that a solitary Proca field requires a background
obeying source-free Maxwell equations, whereas an arbitrary HS field requires that the EM field strength satisfy
the condition $\de_{(\m}F_{\n)\r}=0$.
We also find that an isolated massive HS particle may have consistent non-minimal local interactions with
a gravitational background, which (a) has no restrictions for $s=1$, (b) must be a Ricci symmetric space for $s=2$,
and (c) must be a Ricci symmetric space with additional conditions for $s\geq3$ on the Weyl tensor:
$\nb_{(\m}W_\n{}^\a{}_{\r)}{}^\b=0$. Curiously, the consistency of the Lagrangian dynamics of spinning
particles in various dimensions imposes similar restrictions on the backgrounds~\cite{Spinning,Andrew}. Also, the
study of gravitating partially massless spin-2 fields leads naturally to Ricci symmetric spaces~\cite{ADJ}, which
however do not suffice for consistency in this case.

Let us emphasize that our results for the non-minimal couplings show up as consistent deformations on top
of minimal coupling, whose existence has been implicitly assumed. For EM interactions this assumption may not hold;
indeed, massive HS fields may not have an EM charge but still possess higher multipoles. In this case our conclusions
will not be valid. For gravitational interactions, however, our assumption is well justified since minimal coupling
is required in order for the principle of equivalence to hold~\cite{WP}.

A couple of comments on our consistent local deformations are due. First, the choices (\ref{choiceEM})
and~(\ref{choiceGR}) for the first order corrections respectively for EM and gravity are unique in that they
not only keep locality intact, but also make the inclusion of higher order deformations unnecessary. In
particular, they leave the covariant transversality conditions undeformed. Any other choices for these first order
deformations will call for either non-locality or higher order corrections or both. Last but not the least,
the restrictions on the EM and gravitational backgrounds are required only for the sake of locality. This
does not mean that a massive HS field cannot propagate consistently in an arbitrary background, but that it
can do so only if non-local interactions are allowed and/or other interacting DoFs are present. It is possible
that if one begins with an enlarged (possibly infinite) set of HS fields, one may find consistent local
deformations of the free theory for arbitrary backgrounds. While from bottom-up it is a priori not possible
to know which set of fields, if any, may achieve this feat, top-down approaches like KK incarnations of string
theory or supergravity provide some concrete examples of this kind.

Our results cover a wider range of possibilities since the approach does not assume any Lagrangian embedding.
Requiring that the system of equations results from a local Lagrangian may therefore result in
further restrictions. Indeed, Lagrangians for a charged massive HS field are known to exist for a constant EM
background only in $d=26$~\cite{AN2,PRS}; it is not clear whether they exist in arbitrary dimensions,
which may require additional dynamical fields~\cite{2plus0}. Another example is a massive spin-2 field in a
gravitational background; a local Lagrangian approach seem to call for Einstein manifolds~\cite{Buchbinder}$-$a
subset of Ricci symmetric manifolds we require. In short, the requirements we find are necessary
but may not be sufficient for the existence of local Lagrangians.

The possible non-existence of a local Lagrangian is perhaps the most uncomfortable feature of this formalism.
How does one quantize such non-Lagrangian involutive systems? The authors in Refs.~\cite{KaLySh,KapLySh} have
tried to address this issue through a formalism based on a generalized Lagrangian structure, called the Lagrange
anchor. This structure contains information beyond the solutions to the EoMs, much like the Lagrangian does, that
can help one perform a path integral for the non-Lagrangian system.

The approach of deformation of involutive equations is a powerful one that can be applied to many different
systems. The immediate things to consider are fermions and mixed symmetry tensors in EM and gravitational
backgrounds~\cite{CRS}. One may also try partially massless fields to see if the results of~\cite{ADJ} could be
reproduced. Note that the Lagrangian formulation requires auxiliary fields, which can be incorporated into the
trace of some
otherwise traceless tensor fields. In order to make more contact with the Lagrangian framework, it is desirable
to start with traceful fields. In this case, as one casts the original system into an involutive form, the free
system enjoys a much richer gauge structure with many more gauge identities~\cite{CRS}. The correct DoF count
crucially depends on taking all the independent identities into account. Unlike in the traceless set-up, there
exist relations among the gauge identities themselves; one needs to be careful while deforming the free system.
We leave this as future work.

\section*{Acknowledgments}

RR and MS would like to thank S.~L.~Lyakhovich and A.~Waldron for stimulating discussions and useful comments,
the Galileo Galilei Institute for Theoretical Physics (where this work was initiated) for hospitality and the
INFN for partial support.
IC acknowledges partial support by CONACyT's program ``Estancias Sab\'aticas y Posdoctorales al Extranjero''
(M\'exico) 2011-2012, grant number 172676. RR is a Postdoctoral Fellow of the Fonds de la Recherche Scientifique-FNRS.
The work of IC and RR is partially supported by IISN-Belgium (conventions 4.4511.06 and 4.4514.08) and by the
``Communaut\'e Fran\c{c}aise de Belgique'' through the ARC program and by the ERC through the ``SyDuGraM'' Advanced Grant.
MS thanks M.~Henneaux and H.~Nicolai for hospitality respectively at Universit\'e Libre de Bruxelles and at AEI for
Gravitational Physics (Potsdam), where part of his work was done. He acknowledges support from DST (India) in the form of
a project.

\end{document}